\documentclass[a4paper,11pt,dvips]{article}
\usepackage{graphicx}
\usepackage{amsmath}
\usepackage{amssymb}
\usepackage{latexsym}

\newcommand{\bra}{\begin{array}}
\newcommand{\era}{\end{array}}
\newcommand{\beq}{\begin{equation}}
\newcommand{\eeq}{\end{equation}}
\newcommand{\beqar}{\begin{eqnarray}}
\newcommand{\eeqar}{\end{eqnarray}}

\def\BC{\bb C}
\def\_\BC{\bbi C}

\def\( {\left(}
   \def\) {\right)}
\def\[ {\left[}
\def\] {\right]}
\def\no2 {{\textstyle{n\over 2}}}

\newcommand{\be}{\beta}

\newcommand{\al}{\alpha}

\newcommand{\lga}{\longrightarrow}

\begin{document}
\begin{titlepage}
\setcounter{page}{1}
\renewcommand{\thefootnote}{\fnsymbol{footnote}}

\begin{flushright}
ucd-tpg:09.02\\
arXiv:0906.0097
\end{flushright}

\vspace{5mm}
\begin{center}

%\begin{document}
%\thispagestyle{empty}

%\begin{flushright}
%ucd-tpg/06xx\\
%hep-th/yymmxxx
%\end{flushright}

\vspace{0.5cm}
%\begin{center}
 {\Large \bf Tunneling for Dirac Fermions in Constant Magnetic Field}

\vspace{0.5cm}

{\bf El Bou\^azzaoui Choubabi$^a$\footnote{choubabi@gmail.com}, Mohamed El Bouziani$^a$}
and {\bf Ahmed Jellal$^{a,b}$\footnote{jellal@pks.mpg.de, jellal.a@ucd.ac.ma}}\\
\vspace{0.5cm}

$^a${\em Theoretical Physics Group,  %Department of Physics,
Faculty of Sciences, Choua\"ib Doukkali University},\\
{\em  PO Box 20, 24000 El Jadida,
Morocco}

$^b${\em Physics Department, College of Science, King Faisal University,\\
PO Box 380, Alahsa 31982,
Saudi Arabia}

\vspace{3cm}

\begin{abstract}

The tunneling effect of two-dimensional Dirac fermions in
a constant magnetic field is studied.
This can be done by using the continuity equation at some points
 to determine the corresponding
reflexion and transmission coefficients. For this, we consider a system
made of graphene as superposition of two different
regions where the second is characterized by
an energy gap $t'$.
In fact, we treat concrete
systems to practically give two illustrations: barrier
and diode. For each case, we discuss the transmission in terms of the
ratio of the energy conservation and $t'$. Moreover, we analyze
the resonant tunneling by introducing a scalar Lorentz potential
where it is shown that a total transmission is possible.
\end{abstract}
\end{center}
\end{titlepage}

\newpage

%%%%%%%%%%%%%%%%%%%%%%%%%%%%%%%%%%%%%%%%%%%%%%
\section{Introduction}
%%%%%%%%%%%%%%%%%%%%%%%%%%%%%%%%%%%%%%%%%%%%%%%%%

After the experimental realization of graphene in 2005,
this kind of system became an attractive
subject not only for experimentalists
but also for theoretician physicists.
This is because of the nature of its
structures and the behavior of its
relativistic particles. In addition of
 exhibiting the anomalous quantum Hall effect~\cite{novoselov,zhang},
graphene gives an example of condensed matter
physics where  the quantum electrodynamics tools can be
applied~\cite{katsnelson}. These new developments offered
a laboratory for many investigations where interesting results are
obtained by solving different problems. Because of the relativistic nature
of their fermions,
the system made of graphene renewed
the interest of studying the Dirac fermions in two-dimensions.

One of characteristics of Dirac fermions in graphene is their
ability to tunnel through a potential barrier with probability
one~\cite{NaturePhys,Bai07}. This so called Klein tunneling of chiral
particles has long ago been proposed in the framework of quantum
electrodynamics~\cite{Klein29,Calogeracos,Zuber}, but was never
observed experimentally. As appealing as the Klein tunneling may sound from the point of view
of fundamental research, its presence in graphene is unwanted when
it comes to applications of graphene to nanoelectronics. This comes
about because the pinch-off the field effect transistor may be very
ineffective. The same may occur because of the minimum conductivity
of graphene at the neutrality point. One way to
overcome these difficulties is by generating a gap in the spectrum.
From the point of view of Dirac fermions this is equivalent to the
generation of a mass term.

 A possibility of generating gaps in the graphene spectrum is
to deposit graphene on top of hexagonal boron nitride (BN)~\cite{Giovannetti07}.
This material is a band gap insulator with a
boron to nitrogen distance of the order of 1.45 \AA~\cite{Zupan71}
(in graphene the carbon-carbon distance is 1.42 \AA) and a gap of
the order of $4$ {\ttfamily eV}. It was shown that in the most
stable configuration, where a carbon  is on top of a boron and the
other carbon in the unit cell is centered above a BN ring, the value
of the induced gap is of the order of 53 m{\ttfamily eV}. Depositing
graphene on a metal surface with a BN buffer layer leads  to $n-$doped
graphene with an energy gap of 0.5 {\ttfamily eV}~\cite{Lu07}.

Theoretically, the tunneling effect of system type SiO$_2$-BN in zero
magnetic field
is discussed~\cite{peres}. In fact, it is assumed that it is possible
to manufacture slabs with SiO$_2$-BN interfaces, on top of which a graphene
flake is deposit. This will induce spatial regions where graphene has
a vanishing gap intercalated with regions where the BN will cause a
finite gap. The graphene physics is considered in two
different regions:
  the $k-$region, where the graphene sheet is standing on top
of SiO$_2$, and
  a $q-$region, where a mass-like term is present,
caused by BN, inducing an energy gap
of value $2t'$. %(for all numerical purposes we use $t'=$0.1 {\ttfamily eV}).
The effect of chiral electrons in graphene through
a region is studied where the electronic spectrum changes from the usual linear dispersion
to a hyperbolic dispersion, due to the presence of a gap. It is shown that contrary
to the tunneling through a potential barrier, the transmission of electrons is, in this case,
smaller than one for normal incidence.
%they claimed that
%their mechanism may be useful for designing electronic
%devices made of graphene.

Motivated by the reason discussed above and in particular
the investigation made in~\cite{peres},
we would like to reply an interesting question. %, which attracted our attention. This concerns
 the fact that what happens to the tunneling effect of SiO$_2$-BN
in the presence of an external magnetic field $B$. Otherwise, still we have
the same conclusions reached for the case $B=0$ in~\cite{peres} or they will
be affected.
To answer these inquiries,  we study such system in the same conditions
as in~\cite{peres} but taking into account the effect of
the gauge field. This will allow us to deal with some issues and end up with
different conclusions. Moreover, under some conditions we show that
there is possibility for a total transmission.

Note that, actually
there are many works on tunneling of electrons in graphene in the presence of magnetic field
%Note that, there are actually some works on the tunneling effect in terms of
%magnetic fields
e.g.~\cite{demartino, shytov}. However, they did not consider a system type SiO$_2$-BN and also
their magnetic fields are partially applied to the considered systems as in~\cite{demartino}.
The present work gives a novel approach how to deal with the effect
where a magnetic field is applied to whole system with an energy gap $t'$.

More precisely, we consider a system composed of
two different regions, where the second is characterized by
an energy gap $t'$, in a perpendicular  magnetic field.
The energy spectrum solutions are obtained for both regions
in terms of two Landau levels and $t'$.
From the energy conservation $E$, we obtain a set of the energy values that
allows us to discuss the tunneling effect
By inspecting these values, one can end up with three limiting cases,
which they have  interesting consequences on the reflexion and transmission
of the present system.

To be concrete, we give two examples of system made of graphene.
%and determine the reflexion and transmission
%coefficients in terms of $B$. For this, we start by evaluating the eigenvalue
%solution for each regions and determine the corresponding wavefunctions.
In the beginning, we consider a barrier in magnetic field and
 study the tunneling effect. Indeed, from
the continuity equation, we get different solutions,
which allowed us to explicitly determine
the reflexion and transmission coefficients for
different regions. They are used to show that
the probability condition is satisfied, namely
the sum of these coefficients is one.
To characterize the transmission behavior
we give different figures, which underline its properties
in terms of the energy ratio ${E\over t'}$.
Using the three limiting cases of energy
we discuss the possibility to obtain a total
transmission and give different interpretations.

As far as the diode in magnetic field is concerned,
 we discuss the tunneling effect by splitting the whole system in
three regions where the second different from the first and third,
which they are identical. After getting different coefficients, we
use an appropriate definition to show that the sum of reflexion
and transmission is one. Analyzing this under some conditions,
we conclude that
it is possible to obtain a total transmission. We also discuss
a limiting situation where the barriers are described by
a scalar Lorentz potential. Using the boundary conditions, we
derive different coefficients, which lead again to verify the
probability condition and give different discussions about such
potential.

The present paper is organized as follows. In section $2$, we formulate our
problem by
writing the Hamiltonian's describing two regions of our system.
 The eigenvalue problems will
be solved to obtain the energy spectrum
and its eigenspinors. We inspect the eigenvalue solutions from
the energy conservation point of view to
 underline their properties in section $3$. After establishing all
needed materials, we treat the first illustration of our system, i.e.
the barrier, in section $4$. In section 5,
we consider the diode in magnetic field as a second illustration
where an interesting limit will be investigated.
 We conclude and give some perspectives in last section.

%%%%%%%%%%%%%%%%%%%%%%%%%%%%%%%%%%%%%%%%%%%%%
\section{Hamiltonian formalisms}
%%%%%%%%%%%%%%%%%%%%%%%%%%%%%%%%%%%%%%%%%%%%%%%%%

As claimed before, our system is a superposition
of two different regions $(\rm I)$ and
 $(\rm II)$, with  $(\rm II)$ has
an energy gap $t'$. For this, we start by setting the necessary
tools needed to treat each region separately.
These concerns to write the corresponding Hamiltonian's and
determine
their eigenvalue solution as well as the eigenspinors.

Region $(\rm I)$ can be identified to a two-dimensional subsystem of Dirac fermions
where the Hamiltonian
for one massless relativistic fermion in the presence of a perpendicular magnetic field
is
\begin{equation}\label{hamr1}
H^{(\rm I)}=v_{F} \, \vec{\sigma} \,\cdot \vec{\pi}
\end{equation}
 where  $v_{F}\approx10^{6}ms^{-1}$ is the Fermi velocity and
$\vec{\sigma}=\left( \sigma_{x},\sigma_{y}\right)$  are the Pauli matrices
\begin{equation}
 \sigma_{x}=\left( %
\begin{array}{c c}
0 & 1 \\
1 & 0 \\
\end{array}
\right), \qquad \sigma_{y}=\left( %
\begin{array}{c c}
0 & -i \\
i & 0 \\
\end{array}
\right).
\end{equation}
The conjugate momentum is
$\vec{\pi}=\vec{p}+\frac{e}{c}\vec{A}$, with $\vec{A}$ is a gauge field.
Recall that, region $(\rm II)$ is of finite mass due to $t'$, then
its appropriate Hamiltonian
can be written as
\begin{equation}\label{hamr2}
H^{(\rm II)}=v_{F}\ \vec{\sigma}\cdot \vec{\pi}+t'\sigma_{z}.
\end{equation}
Its clear that, the mass term $t'\sigma_{z}$ makes difference between both
the above Hamiltonian's. This will play a crucial role
in the forthcoming analysis.

%%%%%%%%%%%%%%%%%%%%%%%%%%%%%%%%%%%%%%%%%%
\subsection{First region spectrum }
%%%%%%%%%%%%%%%%%%%%%%%%%%%%%%%%%%%%%%%%%%%%%%

To do our task, we start by determining
the energy spectrum and
its eigenspinors for each involved region. In doing so, let us start
by writing the Hamiltonian (\ref{hamr1}) as
\begin{equation}
H^{(\rm I)}
=v_{F}\, \left( \begin{array}{c c}
0 &\pi_{x}-i \pi_{y}\\
\pi_{x}+i \pi_{y}& 0 \\
\end{array} \right).
\end{equation}
Choosing the Landau gauge $\vec{A}=B\left( 0,x,0\right) $,
the momenta components take the form $\pi_{x}=p_{x}$ and
$\pi_{y}=p_{y}+\frac{eB}{c}x$. These can be used to
 maps $H^{(\rm I)}$ into
\begin{equation}\label{hamI}
H^{(\rm I)} =v_{F}\, \left( \begin{array}{c c}
0 &  p_{x} -i p_{y} -i \frac{eB}{c}x\\
p_{x} +ip_{y}+ i\frac{eB}{c}x& 0 \\
\end{array} \right).
\end{equation}

As usual to get the energy solutions of (\ref{hamI}), one can
use the eigenvalue equation for a given spinor
$\phi^{(\rm I)} =\left(
\begin{array}{c}
\varphi_{1} \\
 \varphi_{2}\end{array}\right)$ of $H^{(\rm I)}$.
 This is
\begin{equation}
H^{(\rm I)} \left(
\begin{array}{c}
\varphi_{1} \\
 \varphi_{2}\end{array}\right)= E^{(\rm I)} \left(
\begin{array}{c}
\varphi_{1} \\
\varphi_{2}\end{array}\right).
\end{equation}
It implies the "kinetic balance" relation
\beqar
&& -iv_{F}\left( i p_{x}+ p_{y}+\frac{eB}{c}x \right)\varphi_{2}=E^{(\rm I)}\varphi_{1}
\label{phi1}\\
&& iv_{F}\left(-ip_{x}+p_{y}+\frac{eB}{c}x \right)\varphi_{1}=E^{(\rm I)}\varphi_{2}.
\label{phi2}
\eeqar
To determine these spinor components, we can map for instance
(\ref{phi1}) into (\ref{phi2}) to obtain a Schr\"odinger-like equation
for $\varphi_{2}$. After calculation, we obtain
\begin{equation}\label{eq9}
v_{F}^{2}\, \left[p_{x}^{2}+\left(p_{y}+\frac{eB}{c}x\right)^{2}-
\frac{eB\hbar}{c}\right]\varphi_{2}
=\left(E^{(\rm I)}\right)^{2}\varphi_{2} \equiv h\varphi_{2}.
\end{equation}
It is similar to that of the harmonic oscillator up to some constant term
and thus $ \varphi_{2}$ can be seen as its eigenfunction. To clarify
this statement, let us consider
the Landau Hamiltonian in the same gauge, such as
\beq
H_0= \frac{1}{2 m}\left[p_{x}^{2}+\left(p_{y}+\frac{eB}{c}x\right)^{2}\right].
\eeq
One can easily show that its energy spectrum reads as
\begin{equation}
E^{(\rm I)'}= \hbar \omega_c\left(n +\frac{1}{2}\right), \qquad n=0,1,2,\cdots
\end{equation}
where the cyclotron frequency
$\omega =\frac{eB}{cm}$. In terms of the cylindrical parabolic
functions, The corresponding wavefunctions are given by
\begin{equation}\label{varph2}
\phi(x,y)= D_{n }(Q)\, e^{ik_{y}y}
\end{equation}
 where $k_{y}$ is a wave vector along $y$-direction in region $(\rm I)$.
We have set $Q=\frac{x+x_{0}}{l_{B}}$
and $x_{0}=k_{y} l_{B}^{2}$, with the magnetic length $l_{B}=\sqrt{\frac{c\hbar}{e B}}$.
It is not hard to verify the property
\begin{equation}\label{prop}
 D_{n}(-Q) =  (-1)^{n} D_{n}\left(Q\right)
\end{equation}
which will be used in the forthcoming analysis
and more precisely when we start to investigate the
tunneling effect. Explicitly,  the states $\phi(x,y)$ are
\begin{equation}
 \phi(x,y) =\left(l_B\sqrt{\pi}n!2^{n}\right)^{-\frac{1}{2}}
\exp\left(-\frac{Q^{2}}{2}\right)H_{n}(Q)\, e^{ik_{y}y}
\end{equation}
where the Hermite polynomials $H_{n}(Q)$ are
\beq
H_{n}(Q)=(-1)^{n}\exp(Q^{2}) \frac{d^{n}}{dQ^{n}} \exp(-Q^{2}).
\eeq
Now injecting (\ref{varph2}) in (\ref{eq9}), one can end up with
\begin{equation}
\left[\frac{2v_{F}^{2}\hbar^{2}}{l_{B}^{2}}(n+\frac{1}{2})-
\frac{v_{F}^{2}\hbar^{2}}{l_{B}^{2}}\right]D_{n}(Q)=\left(E^{(\rm I)}\right)^{2}D_{n}(Q)
\end{equation}
which is leading to the form
\begin{equation}
\left[\frac{2v_{F}^{2}\hbar^{2}}{l_{B}^{2}}n \right]D_{n}(Q)=\left(E^{(\rm I)}\right)^{2}D_{n}(Q).
\end{equation}
It is clear that the eigenvalues read as
 \begin{equation}
\left(E^{(\rm I)}_n\right)^{2}=\alpha^{2}n
\end{equation}
where the constant $\al = \frac{2v_{F}^{2}\hbar^{2}}{l_{B}^{2}}$.
According to the interpretation of matter and antimatter, the bottom valley of
the band structure takes the value $-n$ that correspond to negative energies
(anti-matter). For this reason, we write the eigenvalues $H^{(\rm I)}$ as
\begin{equation}
%E_{1}^{2}=\alpha^{2}|n|\Rightarrow
E^{(\rm I)}_n={\sf sgn}(n)\alpha\sqrt{|n|}, \qquad n\in {\mathbb Z}
\end{equation}
where the eigenfunctions are given by
\begin{equation}
\varphi_{2}(x,y)= D_{|n|}\left({x+x_0\over l_B}\right)\, e^{ik_{y}y}.
\end{equation}

To complete our analysis, we need to determine the
 second spinor component, which can be obtained
from
the "kinetic balance" relation. Doing so to obtain
\begin{equation}
\varphi_{1}={1\over {E^{(\rm I)}}}
\left[{-iv_{F} \left( i p_{x}+ p_{y}+\frac{eB}{c}x \right)D_{|n|}}\right]\, e^{ik_{y}y}.
\end{equation}
It leads to the solution
\begin{equation}
\varphi_{1}=-{\sf sgn}(n)iD_{|n|-1}(Q)\, e^{ik_{y}y}.
\end{equation}
Finally, combining all to get the normalized eigenspinors
as
\begin{equation}
\phi^{(\rm I)}_{(n,k_{y})}(x,y)=
\frac{1}{\sqrt{2}}
\left(%
\begin{array}{c}
-siD_{\mid  n \mid  -1}(x+x_{0})\\
D_{\mid  n \mid }(x+x_{0})
\end{array}%
\right)\, e^{ik_{y}y}
\end{equation}
where $s={\sf sgn}(n)$ and the convention
${\sf sgn}(0)=0$ should be taken into account.
The energy spectrum has zero-mode wavefunction,
such as
\begin{equation}
\phi^{\rm (I)}_{(0,k_{y})}(x,y)=\frac{1}{\sqrt{2}}
\left(%
\begin{array}{c}
0\\
D_{0}(x+x_{0})
\end{array}%
\right) \, e^{ik_{y}y}.
\end{equation}
These results are concerning the first region, which is
assimilated to a subsystem of Dirac fermions in constant magnetic field.
Next we will see how the above results will change when we move
to the second region, which is also Dirac fermions but with a mass
term.

%%%%%%%%%%%%%%%%%%%%%%%%%%%%%%%%%%%%%%%%%%
\subsection{Second region spectrum }
%%%%%%%%%%%%%%%%%%%%%%%%%%%%%%%%%%%%%%%%%%%%%%

From the nature of the system under consideration,
we write
the Hamiltonian corresponding to
region $(\rm II)$ in terms of matrix as
%can matrixially be written as
\begin{equation}
H^{(\rm II)}
=v_{F}\, \left( \begin{array}{c c}
0 &\pi_{x}-i \pi_{y}\\
\pi_{x}+i \pi_{y}& 0 \\
\end{array} \right) +\left( \begin{array}{c c}
t' &  0\\
0 & -t' \\
\end{array} \right).
\end{equation}
In the Landau gauge,  we have the form
%\begin{center}
%$\sigma_{z}=\left(
%\begin{array}{cc}
%1 & 0 \\
%0 & -1
 %                  \end{array}\right) $
%\end{center}
\begin{equation}
H^{(\rm II)} =v_{F}\ \left( \begin{array}{c c}
t' &  p_{x} -i p_{y} -i\frac{eB}{c}x\\
p_{x}+ ip_{y}+ i\frac{eB}{c}x & -t' \\
\end{array} \right).
\end{equation}
Note that, the energy gap behaves like a mass term. Certainly
this will affect the above results and lead to interesting
consequences in underlying the basics features of such
system.

We need to derive the energy spectrum and its
eigenspinors. In doing so,
 let us fix $\phi^{(\rm II)}=\left(
\begin{array}{c}
\varphi '_{1} \\
 \varphi '_{2}\end{array}\right)$ as a spinor of $H^{(\rm II)}$
to write
\begin{equation}
H^{(\rm II)} \left(
\begin{array}{c}
\varphi '_{1} \\
 \varphi '_{2}\end{array}\right)= E^{(\rm II)} \left(
\begin{array}{c}
\varphi '_{1} \\
\varphi '_{2}\end{array}\right)
\end{equation}
which implies two relations
\beqar
&&-iv_{F}\ \left( i p_{x}+ p_{y}+\frac{eB}{c}x \right) \varphi '_{2}=
\left(E^{(\rm II)}-t'\right) \varphi '_{1}\label{r2eq1}\\
&& iv_{F}\ \left(-ip_{x}+p_{y}+\frac{eB}{c}x \right)\varphi '_{1}=
\left(E^{(\rm II)}+t'\right) \varphi '_{2}.\label{r2eq2}
\eeqar
These can be treated as we have done before to get one equation
for one component spinor. After injecting (\ref{r2eq1}) in (\ref{r2eq2}),
we obtain a differential equation of second order for $\varphi '_{2}$. This
is
\begin{equation}
h\varphi '_{2} =\left[\left(E^{(\rm II)}\right)^{2}-t'^{2}\right]\varphi '_{2}.
\end{equation}
 It solution gives the second spinor component as
\begin{equation}\label{pp2}
 \varphi '_{2}(x,y)=D_{|m|}\left({x+x'_0\over l_B}\right) \, e^{iq_{y}y},
\qquad m\in {\mathbb Z}
\end{equation}
where $q_{y}$ is a wave vector along $y$-direction in the
second region and $ x_{0}'=q_{y} l_{B}^{2}$. From last equation, it is easy to obtain
the energy spectrum
\begin{equation}
E^{(\rm II)}={\sf sgn}(m)\sqrt{\alpha^{2}|m| +t'^{2} }.
\end{equation}
We notice that the term $t'$ makes difference with respect to spectrum of
region ${(\rm I)}$. It is convenient for our task to write energy
as
\begin{equation}
E^{(\rm II)}-t'=\frac{\alpha^{2}|m|}{ E^{(\rm II)}+t' }.
\end{equation}
%This will play a crucial role in the forthcoming
%analysis. In fact,
This will be used to show how the tunneling effect
behaves in terms of the energy conservation and the parameter $t'$.

Returning to the "kinetic balance" relations (\ref{r2eq1}) and (\ref{r2eq2}),
% using the same analysis as
%we have done before,
one can easily derive the first spinor component
$\varphi '_{1}$. It is
\begin{equation}
 \varphi '_{1}(x,y)=\frac{-i\alpha\sqrt{|m|}}{E^{(\rm II)}-t'}
D_{|m|-1}\left({x+x'_0\over l_B}\right) \, e^{iq_{y}y}.
\end{equation}
After normalization the eigenspinors read as
\begin{equation}
\phi^{\rm (II)}_{(m,q_{y})}(x,y)=\frac{1}{\sqrt{2}}\left(%
\begin{array}{c}
-a_{m}i D_{\mid m \mid -1}(x+x_{0}')\\
b_{m}D_{\mid  m \mid }(x+x_{0}')
\end{array}%
\right)\, e^{iq_{y}y}.
\end{equation}
The normalization constants are given by
\beq
a_{m} =s'\sqrt{\frac{E^{(\rm II)}+s't'}{E^{(\rm II)}}}, \qquad
b_{m} =\sqrt{\frac{E^{(\rm II)}-s't'}{E^{(\rm II)}}}
\eeq
where we have set $s'= {\sf sgn}(m)$. They verify the useful relation
\beq
a_m^2 + b_m^2 = 2.
\eeq
We have also here a zero-mode wavefunction
\begin{equation}
\phi^{\rm (II)}_{(0,k_{y})}(x,y)=\frac{1}{\sqrt{2}}
\left(%
\begin{array}{c}
0\\
D_{0}(x+x'_{0})
\end{array}%
\right)\, e^{iq_{y}y}.
\end{equation}
 Clearly, in the interface between
two regions, there is conservation of the tangent  components of
the wave vector, i.e.  $k_{y}$ = $q_{y}$,
which is equivalent to $x_0=x'_0$. This conclude the
investigation of the energy spectrum for the present system
in a constant magnetic field. Before using these
results to solve some issues, it is interesting
to start by underling
properties of the eigenvalues.

%%%%%%%%%%%%%%%%%%%%%%%%%%%%%%%%%%%%%%%%%%%%%%
\section{Selection rules}
%%%%%%%%%%%%%%%%%%%%%%%%%%%%%%%%%%%%%%%%%%%%%%%%%

Since we are considering two different regions
characterized by two quantum numbers,
one may investigate their behaviors
in terms of the energy system and the parameter
$t'$. For this, we use the energy conservation
to establish a relation between them,
which will be employed in the next in dealing with
different issues.  In particular, it will used
to
 discuss  the tunneling effect for our system
through the reflexion and transmission coefficients.

Recall that, from the above analysis, we ended up with two energy
spectra: $E^{\rm (I)}$ and  $E^{\rm (II)}$. On the other hand,
due to the system nature one should have an energy conservation, such as
\beq\label{encons}
\left(E^{\rm (I)}\right)^2 =\left(E^{\rm (II)}\right)^2=E^{2}.
\eeq
After replacing the energies by their expressions, it is easy to observe
that the allowed energy values should verify the relation
\begin{equation}\label{eq02}
 \frac{E^{2}}{t'^{2}}=\frac{|n|}{|n|-|m|}
\end{equation}
where the constraint $|m|\leq|n|$ must be fulfilled.
According to (\ref{eq02}),
one can realize $n$ and $m$ in terms of a pair of quantum numbers $(p,q)$,
such as
\begin{equation}
n=\pm kp, \qquad m=\pm k(p-q)
\end{equation}
where $\frac{p}{q}$ is  an irreducible fraction %whose $p$ and $q$ are integers first values,
with $p\geq q$ and $k$ is an integer value. This realization
can be written in compact form as
$(n,m)=\pm k(p,p-q)$,
%\end{equation}
which is equivalent to say that this pair lies in the set
of values
\beq
(n,m)= \left\{ \pm (p,p-q),\pm 2(p,p-q),\pm 3 (p,p-q), \pm 4(p,p-q), \cdots\right\}.
\eeq
This is very important because
 without such set one can not
talk about tunneling effect in the present case.
We will clarify
this statement from next section and exactly
when we start to calculate different quantities in order to
check the probability condition.

The relation (\ref{eq02}) is important in sense that its has some interesting
consequences and can be used to give different interpretations of
the present system.
%, we notice that there are three
%interesting cases, which fix the system behavior.
%It is relevant to discuss some interesting limits those will have some consequences
%in the forthcoming analysis.
In fact,
by inspecting some limiting cases, one can show
\begin{itemize}
\item $ n=0 \Longrightarrow m =0$.
\item $ \frac{E^{2}}{t'^{2}} \lga \infty \Longrightarrow n=\pm m
\Longrightarrow a_{m}=s'=s $,  $b_{m}= 1$.
 \item $\frac{E^{2}}{t'^{2}}=1 \Longrightarrow m=0 $.
\end{itemize}
In the next, we will see how these cases affect the tunneling effect
through the evaluation of the reflexion and transmission coefficients.
%More precisely, they will fix the nature of the involved system.

On the other hand, one can plot different quantum numbers to
underline their behaviors. For this, we present three figures (1,2,3) illustrating
some cases where the bold dots are exactly the allowed values for
each numbers. They are
\vspace{3mm}
%%%%%%%%%%%%%%%%%%%%%%%%%%%%%%%%%%%%%%%%%%%%%%%%%%%%%%%%%%%%
%\begin{figure}
%\begin{figure}[ht]
%\begin{figure}[!t]
\begin{center}
\includegraphics [width=12.17cm,keepaspectratio]{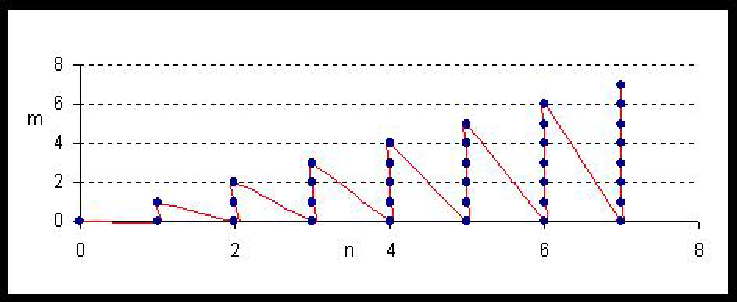}
\end{center}
%\end{figure}
%\caption
%\begin{center}
{Figure 1: Variation of  $m$ in terms of $n$, it is showing that
for $n=5$ there are 6 possible values for $m$, i.e.
$m=0,1,\cdots 5$.}
%\label{fig2}
%\end{center}
%\end{figure}
%%%%%%%%%%%%%%%%%%%%%%%%%%%%%%%%%%%%%%%%%%%%%%%%%%%%%%%
\vspace{3mm}

%%%%%%%%%%%%%%%%%%%%%%%%%%%%%%%%%%%%%%%%%%%%%%%%%%%%%%%%%
%\begin{figure}
\begin{center}
\includegraphics [width=12.17cm,keepaspectratio]
{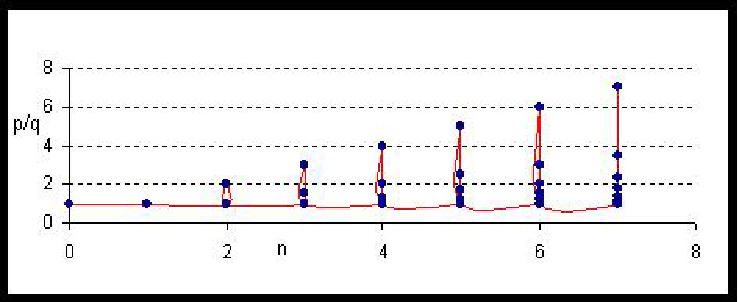}
 % Image1.jpg: 354x155 pixel, 96dpi, 9.37x4.10 cm, bb=0 0 266 116
%\caption
\end{center}
%\begin{center}
{Figure 2:
Variation of the ratio of energy conservation and $t'$, $\frac{p}{q}$, in terms of $n$.
For $n=0$, we have $\frac{p}{q}=1$, which is in agreement with the first item cited above.}
%\end{center}
%\end{figure}
%%%%%%%%%%%%%%%%%%%%%%%%%%%%%%%%%%%%%%%%%%%%%%%%%%%%%%%%
\vspace{3mm}
%%%%%%%%%%%%%%%%%%%%%%%%%%%%%%%%%%%%%%%%%%%%%%%%%%%%%%%%%%%%%%%%%%%
%\begin{figure}
\begin{center}
\includegraphics [width=12.17cm,keepaspectratio]
{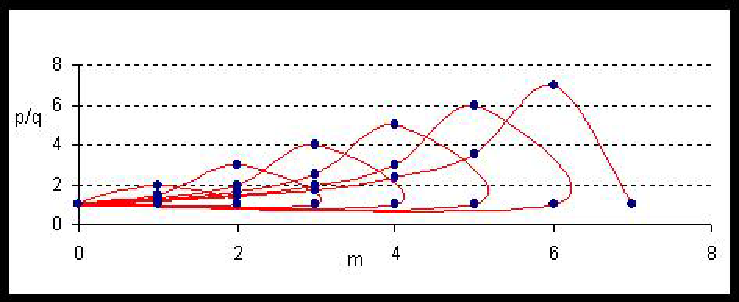}
 % Image1.jpg: 354x155 pixel, 96dpi, 9.37x4.10 cm, bb=0 0 266 116
%\caption
\end{center}
%\begin{center}
{Figure 3: Variation of the ratio of energy conservation and $t'$, $\frac{p}{q}$,
 in terms of $m$. For $m=0$, we have identical behavior as for $n=0$.}
%\end{center}
%\end{figure}
%%%%%%%%%%%%%%%%%%%%%%%%%%%%%%%%%%%%%%%%%%%%%%%%%%%%%%%%%%%%%%%%%%%%%%

%%%%%%%%%%%%%%%%%%%%%%%%%%%%%%%%%%%%%%%%%%%%%%%%%%%%%%%%%%%%%
\section{ Barrier in magnetic field} %Transmission and reflexion between two regions}
%%%%%%%%%%%%%%%%%%%%%%%%%%%%%%%%%%%%%%%%%%%%%%%%%%%%%%%%%%%%%%

To treat a concrete example of the present system, we consider the barrier
in magnetic field. This latter can be seen as superposition of two regions
separated by interface. We study the  tunneling effect by
evaluating the reflexion and transmission coefficients at
interface, which corresponds to the point zero. These
will be used to show that the probability condition is
exactly one and emphasis what makes difference with respect
to the same system studied in zero field~\cite{peres}. To do this task, we distinguish
two cases: propagation with positive and negative incidences.
In both cases we deal with propagation from region $\rm(I)$
to $\rm(II)$ and vice verse.

%%%%%%%%%%%%%%%%%%%%%%%%%%%%%%%%%%%%%%%%%%%%%%%%%%%%%%%%%%%
\subsection{Propagation with positive incidence}
%%%%%%%%%%%%%%%%%%%%%%%%%%%%%%%%%%%%%%%%%%%%%%%%%%%%%%%%%%%

We start by defining what does mean propagation with positive incidence
in our case. In fact, with this we will be able to do our job
and therefore use the continuity equation at interface
to determine different quantities.
%
%our analysis by considering  propagation with positive incidence
%and explicitly determine the reflexion and transmission coefficients.
Indeed, we split the obtained eigenspinors for two regions
in positive and negative directions of the variable $x$,
which is equivalent to write in region ${\rm (I)}$
\beq\label{firan}
\phi^{\rm (I)}_{+}=\frac{1}{\sqrt{2}}\left(
\begin{array}{c}
-si D_{|n|  -1}(x+x_{0})\\
 D_{|n| }(x+x_{0})
 \end{array}
\right)\,  e^{ik_{y}y}, \qquad
\phi^{\rm (I)}_{-} =\frac{1}{\sqrt{2}}\left(
\begin{array}{c}
-si D_{|n|  -1}(-x-x_{0})\\
 D_{|n| }(-x-x_{0})
                \end{array}\right) \,  e^{ik_{y}y}
\eeq
and in ${\rm (II)}$ we can do the same, such as
\beq\label{seran}
\phi^{\rm (II)}_{+} = \frac{1}{\sqrt{2}}\left(
\begin{array}{c}
-a_{m} i D_{|m| -1}(x+x_{0})\\
b_{m}D_{|m| }(x+x_{0})
\end{array} \right) \,  e^{ik_{y}y}, \qquad
\phi^{\rm (II)}_{-} = \frac{1}{\sqrt{2}}\left(
\begin{array}{c}
-a_{m}i D_{|m| -1}(-x-x_{0})\\
b_{m}D_{|m| }(-x-x_{0})
\end{array} \right) \,  e^{ik_{y}y}
\eeq
where we used the labels $(\pm)$ to
show which direction is concerned.
We will see how
these materials will serve as a tool to
deal with different issues. For this, one has
to consider two different cases, which
will be the next task.

%%%%%%%%%%%%%%%%%%%%%%%%%%%%%%%%%%%%%%%%%%%%%%%%%%%%%%%%%%%%%%%%%%%%
{\bf Propagation from region ${\rm (I)}$ to region ${\rm (II)}$:}
%%%%%%%%%%%%%%%%%%%%%%%%%%%%%%%%%%%%%%%%%%%%%%%%%%%%%%%%%%%%%%%%%%%%
This means that the waves are propagating from left to right
and due to the continuity of the system, one can establish
two equations. Indeed,
at the interface $x=0$ and for all $y$, we can write
\begin{equation}\label{reftr}
 \phi^{\rm (I)}_{+}+r_{nm}^{+}\phi^{\rm (I)}_{-}=t_{nm}^{+}\phi^{\rm (II)}_{+}
\end{equation}
where $r_{nm}^{+}$ and $t_{nm}^{+}$ are reflexion and transmission coefficients,
respectively,  $(+)$ indicates
sign of propagation. From  (\ref{firan}) and (\ref{seran}), we end up with
\beq
\left(
\begin{array}{c}
-s iD_{|n|-1}(x_{0})\\
D_{|n|}(x_{0})
                \end{array}\right) +r_{nm}^{+}\left(
\begin{array}{c}
-s iD_{|n|-1}(-x_{0})\\
 D_{|n|}(-x_{0})
                \end{array}\right)
=t_{nm}^{+}\left(\begin{array}{c}-a_{m} iD_{|m|-1}(x_{0})\\
b_{m}D_{|m|}(x_{0})\end{array}\right).
\eeq
Using the property (\ref{prop}), it is not hard to obtain
\beqar %\label{sys}
&&r_{nm}^{+} (-1)^{|n|}sD_{|n|-1}(x_{0})+t_{nm}^{+}
a_{m}D_{|m|-1}(x_{0}) =s D_{|n|-1}(x_{0})\label{sys}\\
&& -r_{nm}^{+} (-1)^{|n|}D_{|n|}(x_{0})+t_{nm}^{+}b_{m} D_{|m|}(x_{0}) = D_{|n|}(x_{0}).\label{sys2}
\eeqar
They can be solved to explicitly get  the coefficients in terms of some
constants, which are magnetic field dependent. They are
\beqar
r_{nm}^{+} &=&\frac{(-1)^{|n|}\left[sb_{m}A_{nm}(x_{0})-a_{m}B_{nm}(x_{0})\right]}
{sb_{m}A_{nm}(x_{0})+a_{m}B_{nm}(x_{0})} \\
t_{nm}^{+} &=&\frac{2sC_{n}(x_{0})}{sb_{m}A_{nm}(x_{0})+a_{m}B_{nm}(x_{0})}
\eeqar
where different involved quantities are given by
\beqar
A_{nm}(x_{0})&=&D_{|n|-1}(x_{0})D_{|m|}(x_{0}),\qquad
B_{nm}(x_{0}) = D_{|m|-1}(x_{0})D_{|n|}(x_{0}),\nonumber \\
 C_n(x_{0})&=& D_{|n|-1}(x_{0})D_{|n|}(x_{0}).
\eeqar
These coefficients are just part of a set of quantities
that should be completely determined. In fact, we still need
other quantities to achieve our task, which
can be obtained from next consideration.

%%%%%%%%%%%%%%%%%%%%%%%%%%%%%%%%%%%%%%%%%%%%%%%%%%%%%%%%%%%%%%%%%%%%%%%
{\bf Propagation from region ${\rm (II)}$ to region ${\rm (I)}$:}
%%%%%%%%%%%%%%%%%%%%%%%%%%%%%%%%%%%%%%%%%%%%%%%%%%%%%%%%%%%%%%%%%%%%%%
In similar way to the former case, we use the continuity at
point zero to write
\beq\label{ceni}
\phi^{\rm (II)}_{+}+r_{mn}^{+}\phi^{\rm (II)}_{-}=t_{mn}^{+}\phi^{\rm (I)}_{+}.
\eeq
Comparing this with (\ref{reftr}), we notice that there is an index
interchange between different coefficients. By replacing,
(\ref{ceni}) becomes
\beq
\left(
\begin{array}{c}
-a_{m} iD_{|m|-1}(x_{0})\\
b_{m}D_{|m|}(x_{0})
                \end{array}\right) +r_{mn}^{+}\left(
\begin{array}{c}
-a_{m} iD_{|m|-1}(-x_{0})\\
 b_{m}D_{|m|}(-x_{0})
                \end{array}\right)
=t_{mn}^{+}\left(\begin{array}{c}- isD_{|n|-1}(x_{0})\\
D_{|n|}(x_{0})\end{array}\right).
\eeq
These lead to the reflexion and transmission coefficients
\beqar
r_{mn}^{+}&=&\frac{(-1)^{|m|}\left[a_{m}B_{nm}(x_{0})-sb_{m}A_{nm}(x_{0})\right]}
{sb_{m}A_{nm}(x_{0})+a_{m}B_{nm}(x_{0})}
\\
t_{mn}^{+}&=&\frac{2a_{m}F_{m}(x_{0})}{sb_{m}A_{nm}(x_{0})+a_{m}B_{nm}(x_{0})}
\eeqar
where $F_{m}$ is given by
\beq
F_{m}(x_{0})=D_{|m|-1}(x_{0})D_{|m|}(x_{0}).
\eeq
This summarizes our analysis for propagation with positive incidence,
which together will be used to discuss different issues
and before doing so, we need
to analyze negative incidence.

%%%%%%%%%%%%%%%%%%%%%%%%%%%%%%%%%%%%%%%%%%%%%%%%%%%%%%
\subsection{Propagation with negative incidence}
%%%%%%%%%%%%%%%%%%%%%%%%%%%%%%%%%%%%%%%%%%%%%%%%%%%%%%%

One may ask about propagation with negative incidence.
To reply this inquiry, we use the same analysis as before
but one should take into account the negative sign of
variable. For the neediness, we write
the corresponding eigenspinors as
\beq\label{niwf+}
\phi^{\rm(I)}_{+} =  \frac{1}{\sqrt{2}}\left(
\begin{array}{c}
-si D_{\mid  n \mid  -1}(-x-x_{0}) \\
 D_{\mid  n \mid }(-x-x_{0})
                \end{array}\right) \, e^{-ik_{y}y},\qquad
\phi^{\rm(I)}_{-}  = \frac{1}{\sqrt{2}}\left(
\begin{array}{c}
-si D_{\mid  n \mid  -1}(x+x_{0})\\
 D_{\mid  n \mid }(x+x_{0})
                \end{array}\right)\, e^{-ik_{y}y}
\eeq
and for second region, we have
\beq\label{niwf-}
\phi^{\rm(II)}_{+}  = \frac{1}{\sqrt{2}}\left(
\begin{array}{c}
-a_{m} iD_{\mid m \mid -1}(-x-x_{0})\\
b_{m}D_{\mid  m \mid }(-x-x_{0})
\end{array} \right) \, e^{-ik_{y}y},\qquad
 \phi^{\rm(II)}_{-}  = \frac{1}{\sqrt{2}}\left(\begin{array}{c}
-a_{m} iD_{\mid m \mid -1}(x+x_{0})\\
b_{m}D_{\mid  m \mid }(x+x_{0})
\end{array} \right) \, e^{-ik_{y}y}.
\eeq
By analogy to the former case, we distinguish the sense of propagation and determine
the corresponding coefficients in terms of the magnetic field.

%%%%%%%%%%%%%%%%%%%%%%%%%%%%%%%%%%%%%%%%%%%%%%%%%%%%%%%%%%%%%%%%%%%%
{\bf Propagation from region ${\rm (I)}$ to region ${\rm (II)}$:}
%%%%%%%%%%%%%%%%%%%%%%%%%%%%%%%%%%%%%%%%%%%%%%%%%%%%%%%%%%%%%%%%%%%%
Let us start from the continuity equation at point zero, such as
\beq\label{nice}
\phi^{\rm(I)}_{+} +r_{nm}^{-} \phi^{\rm(I)}_{-} =t_{nm}^{-} \phi^{\rm(II)}_{+}
\eeq
where label $(-)$ carried by the involved coefficients describes
sense of propagation. Using (\ref{niwf+}) and (\ref{niwf-})
to rewrite (\ref{nice}) as
\beq
\left(
\begin{array}{c}
-s iD_{|n|-1}(-x_{0})\\
D_{|n|}(-x_{0})
                \end{array}\right) +r_{nm}^{-}\left(
\begin{array}{c}
-s iD_{|n|-1}(x_{0})\\
 D_{|n|}(x_{0})
                \end{array}\right)
=t_{nm}^{-}\left(\begin{array}{c}- a_{m}iD_{|m|-1}(-x_{0})\\
b_{m}D_{|m|}(-x_{0})\end{array}\right).
\eeq
It can be solved to get
\beqar
r_{nm}^{-} &=&\frac{(-1)^{|n|}\left[sb_{m}A_{nm}(x_{0})-a_{m}B_{nm}(x_{0})\right]}
{sb_{m}A_{nm}(x_{0})+a_{m}B_{nm}(x_{0})}
\\
t_{nm}^{-} &=& \frac{(-1)^{|n|+|m|}2sC_{n}(x_{0})}{sb_{m}A_{nm}(x_{0})+a_{m}B_{nm}(x_{0})}.
\eeqar

%%%%%%%%%%%%%%%%%%%%%%%%%%%%%%%%%%%%%%%%%%%%%%%%%%%%%%%%%%%%%
{\bf Propagation from ${\rm (II)}$ to region ${\rm (I)}$:}
%%%%%%%%%%%%%%%%%%%%%%%%%%%%%%%%%%%%%%%%%%%%%%%%%%%%%%%%%%%%%
In similar way, we have at zero point
\beq
\phi^{\rm(II)}_{+} +r_{mn}^{-} \phi^{\rm(II)}_{-} =t_{mn}^{-}\phi^{\rm(I)}_{+}
\eeq
showing the relation
\beq
\left(
\begin{array}{c}
-a_{m} iD_{|m|-1}(-x_{0})\\
b_{m}D_{|m|}(-x_{0})
                \end{array}\right) +r_{mn}^{-}\left(
\begin{array}{c}
-a_{m} iD_{|m|-1}(x_{0})\\
 b_{m}D_{|m|}(x_{0})
                \end{array}\right)
=t_{mn}^{-}\left(\begin{array}{c}- siD_{|n|-1}(-x_{0})\\
D_{|n|}(-x_{0})\end{array}\right).
\eeq
The solutions are given by
\beqar
r_{mn}^{-}&=& \frac{(-1)^{|m|}\left[a_{m}B_{nm}(x_{0})-sb_{m}A_{nm}(x_{0})\right]}
{sb_{m}A_{nm}(x_{0})+a_{m}B_{nm}(x_{0})}
\\
t_{mn}^{-}&=&\frac{(-1)^{|n|+|m|}2 a_{m}F_{m}(x_{0})}
{sb_{m}A_{nm}(x_{0})+a_{m}B_{nm}(x_{0})}.
\eeqar
Finally we derived all materials needed to do our job. In fact,
we use the obtained results to give different discussions
and interpretations.

%%%%%%%%%%%%%%%%%%%%%%%%%%%%%%
\subsection{Discussions }
%%%%%%%%%%%%%%%%%%%%%%%%%%%%%%

We show the relevance of different coefficients
derived so far. In fact, we will
combine all of them to
underline the basic features of the barrier
in constant magnetic field. Certainly, from
the above  analysis one can notice that there
are some relations between coefficients, which
can be interpreted as a kind of symmetry. This will be
helpful in sense that one can check the probability condition
and derive other results.

For the reason of do not repeating  equations,
let us choose a pair of index, such as $ i \neq j \in \left\lbrace n,m \right\rbrace $,
to write them in compact forms. Indeed,
from the above results, one can find
\beq
r_{ij}^{\pm}(x_{0}) =  r_{ij}^{\mp}(x_{0}), \qquad
\frac{t_{ij}^{\pm}(x_{0})}{t_{ij}^{\mp}(x_{0})} =
-\frac{r_{ij}^{\pm}(x_{0})}{r_{ji}^{\pm}(x_{0})}=(-1)^{|n|+|m|}.
\eeq
To show the usefulness of these relations, let us define two
coefficients in terms of the above quantities. The first
one is
\beq\label{rho}
\rho(x_0)=r_{ij}^{\pm}(x_{0})r_{ij}^{\mp}(x_{0}) =
\frac{\left[sb_{m}A_{nm}(x_{0})-a_{m}B_{nm}(x_{0})\right]^{2}}
{\left[sb_{m}A_{nm}(x_{0})+a_{m}B_{nm}(x_{0})\right]^{2}}.
\eeq
It is not hard to obtain
\beq
r_{ij}^{\pm}(x_{0})r_{ji}^{\mp}(x_{0}) = \rho (x_{0})\frac{t_{ij}^{\mp}(x_{0})}{t_{ij}^{\pm}(x_{0})}=
\rho(x_{0})\frac{r_{ij}^{\pm}(x_{0})}{r_{ji}^{\pm}(x_{0})}=(-1)^{|n|+|m|+1} \rho(x_{0}).
\eeq
In the same way we can write the second coefficient as
\beq\label{tau}
\tau(x_0)=t_{ij}^{\pm}(x_{0})t_{ji}^{\pm}(x_{0}) =\frac{4sb_{m}a_{m}C_{n}(x_{0})F_{m}(x_{0})}
{\left[sb_{m}A_{nm}(x_{0}) +a_{m}B_{nm}(x_{0})\right]^{2}}=
\frac{4sb_{m}a_{m}A_{nm}(x_{0})B_{nm}(x_{0})}
{\left[sb_{m}A_{nm}(x_{0}) +a_{m}B_{nm}(x_{0})\right]^{2}}.
\eeq
This implies that all quantities defined here can be
related as
\beq
C_{n}(x_{0})F_{m}(x_{0}) = A_{nm}(x_{0})B_{nm}(x_{0})
\eeq
which can easily check it from their expressions given before. Using (\ref{rho})
and (\ref{tau}), we show that they verify the condition
\beq
\rho(x_{0}) + \tau(x_{0}) =1.
\eeq
This is among the interesting results derived so far. In fact, it tells us
the transmission of barrier in magnetic field can not
be greater than one, which is analogue to what obtained in zero
field case~\cite{peres}.
%Note that, the obtained results are valid for all $x$.
%On the other hand,
%To show the relevance of
%different quantities, we plot some figures, which give an idea how coefficients
%behave in terms of magnetic field.

At this level, one can inspect the relations (\ref{rho}) and (\ref{tau}) to see how
they behave with respect to three items cited in section 3. As well see soon these will
change the nature of the present system. Indeed,
taking into account such items, we straightforwardly end up
with the results
\begin{itemize}
\item $n=0 \Longrightarrow$ $ \rho(x_{0}) =0$, $ \tau(x_{0})= 1$.
\item $\frac{E^{2}}{t'^{2}}\lga\infty \Longrightarrow$  $ \rho(x_{0}) =0$,  $ \tau(x_{0})= 1$.
 \item $\frac{E^{2}}{t'^{2}}=1 \Longrightarrow$
$\rho (x_{0})=1$, $\tau(x_{0})  = 0$.
\end{itemize}
These three cases have interesting interpretations in optics physics. More precisely,
in the first and second points, the interface between two regions behaves
like a non-reflective diopter, which means that everything is transmitted, i.e.
total transmission.
However, in last point, it looks like a mirror where the reflexion is total.

To underline the basic properties of
the transmission coefficient, we plot three figures in terms of
the energy ratio $E/t'$ where the conditions $s=s'$ and $E/t'\geqslant1$
are taking into account. These figures can be interpreted as follows.
\begin{itemize}
\item
Figure $4$: One can see  that each curve rises to an asymptote,
 which decreases when $B_{nm}/A_{nm}$ increases.
Clearly, for $B_{nm}/A_{nm}=1$ the asymptote goes to $1$,
for $B_{nm}/A_{nm}=4$  to $0.6$ and finally for $B_{nm}/A_{nm}=8$  to $0.4$.
\item
Figure $5$:  Each curve suddenly reached the value of $1$ and
then decreases to an asymptote, which decreases when $B_{nm}/A_{nm}$ decreases.
For $B_{nm}/A_{nm}=1$ the asymptote goes to $1$, for $B_{nm}/A_{nm}=1/4$
 to $0.6$ and for $B_{nm}/A_{nm}=1/8$  to $0.4$.
\item
Figure $6$:  Both curves tend towards the same asymptote for $B_{nm}/A_{nm}=1/2$.
The curve decreases to the asymptote $0.7$ of the upper,
 but for $B_{nm}/A_{nm}=2$ the curve rises to the asymptote of the lower.
This behavior can be generalized to any case where
 $B_{nm}/A_{nm}=l$ and $B_{nm}/A_{nm}=1/l$,
for all positive value $l$, which is in agreement with the statement cited
in item 2 that means that the transmission is total whenever ${E\over t'}\lga\infty$.
\end{itemize}
\vspace{4mm}

%%%%%%%%%%%%%%%%%%%%%%%%%%%%%%%%%%%%%%%%%%%%%%%%%%%%%%%%%%%%
%\begin{figure}
\begin{center}
\includegraphics [width=12.17cm,keepaspectratio]
{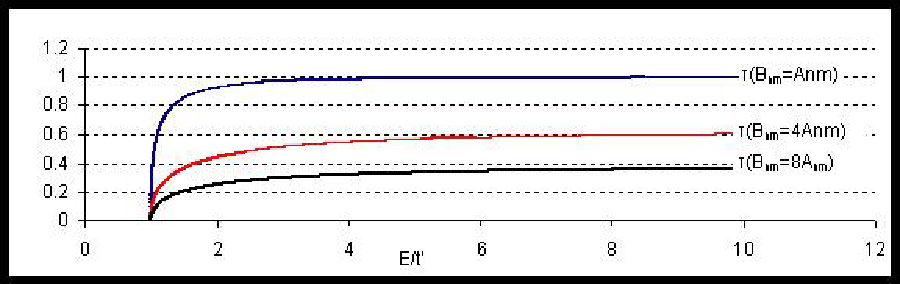}
%\caption
%in terms of $n$}
%\label{fig}
\end{center}
%\end{figure}
%\begin{center}
%\end{center}
%\begin{center}
{Figure 4: Variation of the transmission $\tau (x_0)$ in terms of
the ratio $E/t'$ for three cases:
$B_{nm}=A_{nm}$, $B_{nm}=4A_{nm}$, $B_{nm}=8A_{nm}$}.
%\end{center}
%%%%%%%%%%%%%%%%%%%%%%%%%%%%%%%%%%%%%%%%%%%%%%%%%%%%%%%%%%%%%%%%%%%%%%%%%%%%%%%%%%%%%

\vspace{3mm}
%%%%%%%%%%%%%%%%%%%%%%%%%%%%%%%%%%%%%%%%%%%%%%%%%%%%%%%%%%%%
%\begin{figure}
\begin{center}
\includegraphics [width=12.17cm,keepaspectratio]
{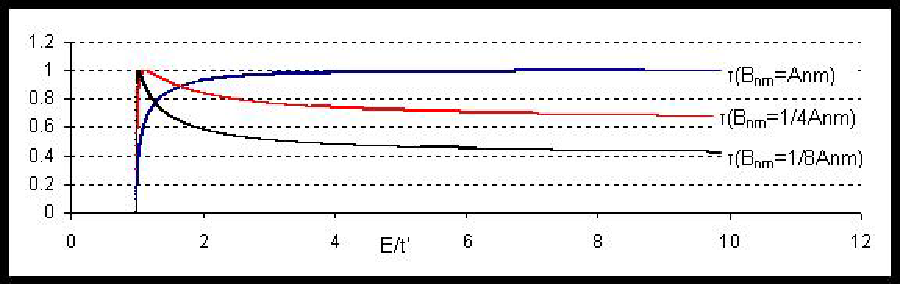}
%\caption
%\label{fig}
\end{center}
%\end{figure}
%%%%%%%%%%%%%%%%%%%%%%%%%%%%%%%%%%%%%%%%%%%%%%%%%%%%%%%
%\begin{center}
{Figure 5: Variation of the transmission $\tau(x_0)$ in
terms of the ratio $E/t'$ but this case for $B_{nm}=A_{nm}$, $B_{nm}=1/4A_{nm}$,
  $B_{nm}=1/8A_{nm}$}.
%\end{center}
%%%%%%%%%%%%%%%%%%%%%%%%%%%%%%%%%%%%%%%%%%%%%%%%%%%%%%%%%%%%%%%%%%%%%%%%%%
%\vspace{3mm}
%%%%%%%%%%%%%%%%%%%%%%%%%%%%%%%%%%%%%%%%%%%%%%%%%%%%%%%%%%%%
%\begin{figure}
\begin{center}
\includegraphics [width=12.17cm,keepaspectratio]
{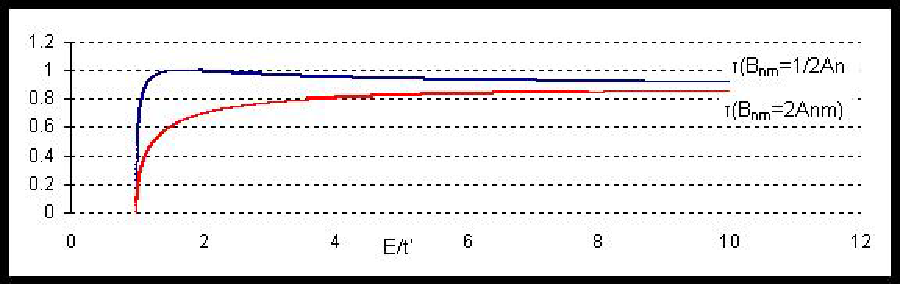}
%\caption
%\label{fig}
\end{center}
%\end{figure}
%\begin{center}
{Figure 6: Variation of the transmission $\tau(x_0)$ in
terms of the ratio $E/t'$ for two cases: $B_{nm}=2A_{nm}$, $B_{nm}=1/2A_{nm}$}.
%\end{center}

%%%%%%%%%%%%%%%%%%%%%%%%%%%%%%%%%%%%%%%%%%%%%%%%%%%%%%%%%%%%%%
\section{Diode in magnetic field }
%%%%%%%%%%%%%%%%%%%%%%%%%%%%%%%%%%%%%%%%%%%%%%%%%%%%%%%%%%%%%%%%%

In last section we focused on the study of two different regions,
with this it is natural to ask about a generalization
to three regions where the second is
characterized also with the energy gap $t'$ and the first and
third are identical. This is the case for instance of a diode
in a constant magnetic field.
To reply this inquiry, let us consider a system composed of
a region indexed by the quantum number $m$ of length $2w$
separating two others
indexed by the same $n$. In fact, we will apply the same machinery as
before to analyze the system behavior in such consideration.

%%%%%%%%%%%%%%%%%%%%%%%%%%%%%%%%%%%%%%%%%%%%%%%%%%%%%%%%
\subsection{Reflexion and transmission coefficients}
%%%%%%%%%%%%%%%%%%%%%%%%%%%%%%%%%%%%%%%%%%%%%%%%%%%%%%%%

The present situation is different from the former one, i.e. the barrier.
This means that actually we have two interfaces fixed at the points
$w_1= x_0 -w$ and $w_2= w+x_0$, which of course will make
difference with respect of the barrier analysis.
We use the above tool to write the continuity equation at each point
and therefore derive the needed quantities for discussing
tunneling effect in such case. In doing so, we consider
also propagation with positive and negative incidences.

In the first interface one can write
\beqar
&&\left(
\begin{array}{c}
-s iD_{|n|-1}(w_1)\\
D_{|n|}(w_1)
                \end{array}\right) +r^{+}\left(
\begin{array}{c}
-s iD_{|n|-1}(-w_1)\\
 D_{|n|}(-w_1)
                \end{array}\right)=\nonumber\\
&&=\alpha \left(
\begin{array}{c}
-a_{m} iD_{|m|-1}(w_1)\\
b_{m}D_{|m|}(w_1)
                \end{array}\right) +\beta \left(
\begin{array}{c}
- a_{m}iD_{|m|-1}(-w_1)\\
 b_{m}D_{|m|}(-w_1)
                \end{array}\right)
\eeqar
where $r^+$, $\alpha$ and $\beta$ are three unknown parameters.
The solution allows us to express two of them in terms of one
for any point $ w_{1}=x_{0}-w$,
such as
\beqar
&& \alpha=\frac{sb_{n}A_{nm}(w_{1}) + a_{m}B_{nm}(w_{1}) + r^{+} (-1)^{|n|}\left[a_{m}B_{nm}(w_{1})-sb_{n}A_{nm}(w_{1})\right]}{2a_{m}b_{m}F_{m}(w_{1})} \\
&& \beta=\frac{a_{m}B_{nm}(w_{1})-sb_{n}A_{nm}(w_{1})   + r^{+} (-1)^{|n|}\left[a_{m}B_{nm}(w_{1})+sb_{n}A_{nm}(w_{1})\right]}
{2(-1)^{|m|}a_{m}b_{m}F_{m}(w_{1})}.
\eeqar
%where $ w_{1}=x_{0}-w$.
These can be simplified to other equations by introducing
two relevant relations in terms of $w_1$.
Indeed, using the former analysis one can show
\beqar
&&\frac{sb_{n}A_{nm}(w_{1}) + a_{m}B_{nm}(w_{1})}{2a_{m}b_{m}F_{m}(w_{1})}=\frac{1}{t^{+}_{mn}(w_{1})}\\
&&\frac{a_{m}B_{nm}(w_{1})+sb_{n}A_{nm}(w_{1}) }{2a_{m}b_{m}F_{m}(w_{1})}=(-1)^{|n|+1}\frac{r^{+}_{nm}(w_{1})}{t^{+}_{mn}(w_{1})}.
\eeqar
With the help of these, we can rewrite $\al$ and $\be$ as
\beqar
&& \alpha= \frac{1}{t^{+}_{mn}(w_{1})}-r^{+}\frac{r^{+}_{nm}(w_{1})}{t^{+}_{mn}(w_{1})}\\
&& \beta=(-1)^{|n|+|m|} \left[\frac{r^{+}}{t^{+}_{mn}(w_{1})} -
\frac{r^{+}_{nm}(w_{1})}{t^{+}_{mn}(w_{1})}\right].
\eeqar
As we will see, they will be employed to set up another couple of equation
and therefore explicitly determine the reflexion and
transmission coefficients for the diode case.

To analyze the second interface, we start by considering
the corresponding continuity equation at the point $w_{2}=x_{0}+w$. This is
\beqar\label{w2eq}
&&\alpha \left(
\begin{array}{c}
a_{m} iD_{|m|-1}(w_2)\\
b_{m}D_{|m|}(w_2)
                \end{array}\right) +\beta (-1)^{|m|}\left(
\begin{array}{c}
- a_{m}iD_{|m|-1}(w_2)\\
b_{m} D_{|m|}(w_2)
                \end{array}\right)=\nonumber\\
&&=t^{+}\left(
\begin{array}{c}
s iD_{|n|-1}(w_2)\\
D_{|n|}(w_2)
                \end{array}\right).
\eeqar
Note that, in obtaining (\ref{w2eq}) in such form we made use
of the property (\ref{prop}).
The coefficients solutions are given by
\beqar
&& \alpha= t^{+}\, \frac{sb_{m}A_{nm}(w_{2}) + a_{m}B_{nm}(w_{2})}{2a_{m}b_{m}F_{m}(w_{2})}=\frac{t^{+}}{t^{+}_{mn}(w_{2})}\\
&& \beta=t^{+}\, (-1)^{|m|}\frac{ a_{m}B_{nm}(w_{2})-sb_{m}A_{nm}(w_{2})}{2a_{m}b_{m} F_{m}(w_{2})}=t^{+}\, (-1)^{|n|+|m|}\frac{r^{+}_{nm}(w_{2})}{t^{+}_{mn}(w_{2})}.
\eeqar
These lead to obtain
 \beqar
r^{+}=\frac{r^{+}_{nm}(w_{1})-r^{+}_{nm}(w_{2})}{1-r^{+}_{nm}(w_{1})r^{+}_{nm}(w_{2})}.
\eeqar
From the above relations, one can find
\beqar
t^{+}=\frac{t^{+}_{mn}(w_{2})}{t^{+}_{mn}(w_{1})}\, \left[\frac{1-r^{+}_{nm}(w_{1})r^{+}_{nm}(w_{1})}{1-r^{+}_{nm}(w_{1})r^{+}_{nm}(w_{2})}\right].
\eeqar
It convenient to write $t^{+}$ in terms of $\rho$ and $\tau$ as
\beqar
t^{+}=\frac{t^{+}_{mn}(w_{2})}{t^{+}_{mn}(w_{1})}\, \left[\frac{1-\rho(w_{1})}{1-r^{+}_{nm}(w_{1})r^{+}_{nm}(w_{2})}\right]=
\frac{t^{+}_{mn}(w_{2})}{t^{+}_{mn}(w_{1})}\,
\left[\frac{\tau(w_{1})}{1-r^{+}_{nm}(w_{1})r^{+}_{nm}(w_{2})}\right].
\eeqar
We close this part by nothing that the obtained results are
only valid for the positive incidence. On the other hand, we
will return to discuss the choice of fixing the points $w_1$ and $w_2$
in terms of $w$.

As far as the negative incidence is concerned, we can use a similar analysis to
derive the corresponding coefficients. Indeed, from above we obtain
\beqar
r^{-}=\frac{r^{-}_{nm}(w_{1})-r^{-}_{nm}(w_{2})}{1-r^{-}_{nm}(w_{1})r^{-}_{nm}(w_{2})}=
\frac{r^{+}_{nm}(w_{1})-r^{+}_{nm}(w_{2})}{1-r^{+}_{nm}(w_{1})r^{+}_{nm}(w_{2})}
\eeqar
as well as
\beqar
t^{-}=\frac{t^{-}_{nm}(w_{2})}{t^{-}_{nm}(w_{1})}\,
\left[\frac{\tau(w_{1})}{1-r^{-}_{nm}(w_{1})r^{-}_{nm}(w_{2})}\right]
=\frac{t^{-}_{nm}(w_{2})}{t^{-}_{nm}(w_{1})}\,
\left[\frac{\tau(w_{1})}{1-r^{+}_{nm}(w_{1})r^{+}_{nm}(w_{2})}\right].
\eeqar
These results will used to deal with the same issues as we have
done for the barrier in magnetic field and also discuss other points.

%%%%%%%%%%%%%%%%%%%%%%%%%%%%%%%%%%%%%%%%%%%%%%%%%%%%%%%%
\subsection{Collecting results}
%%%%%%%%%%%%%%%%%%%%%%%%%%%%%%%%%%%%%%%%%%%%%%%%%%%%%%%%

After deriving the needed tools, we start by discussing
the usefulness of them. For example, we can check the probability
to see how the present system behaves at some critical points. In doing so, let us make
an appropriate definition of the reflexion and transmission
coefficients. It is
 convenient for our task to write
\beq
R=r^{+}r^{-}=r^{2}, \qquad T=t^{+}t^{-}.
\eeq
Explicitly, they take the forms
\beqar
R &=&\frac{\rho(w_{1})+\rho(w_{2})-2r^{+}_{nm}(w_{1})r^{+}_{nm}(w_{2})}
{1+\rho(w_{1})\rho(w^{2})-2r^{+}_{nm}(w_{1})r^{+}_{nm}(w_{2})}\\
T&=&\frac{1-\rho(w_{1})-\rho(w_{2})+\rho(w_{1})\rho(w_{2})}{1+\rho(w_{1})
\rho(w_{2})-2r^{+}_{nm}(w_{1})r^{+}_{nm}(w_{2})}.
\eeqar
These show that how $R$ and $T$ for diode can be linked
to the barrier coefficients at $w_1$ and $w_2$.
Combining all to end up with probability one
\beq
T+R=1.
\eeq
This is among the interesting conclusion reached in this section,
which analogue to that has been obtained
for diode in zero field~\cite{peres}.

One can see how the obtained results for the diode case will change
when we consider our system centered around the point $x_0$.
In such case, its is not hard to show
\beq
r_{nm}^{+}(w_1)=r_{nm}^{+}(w_2).
\eeq
After injecting this relation in the forms of different coefficient
derived here, one can
conclude that everything is transmitted, i.e.
the transmission is total.

%%%%%%%%%%%%%%%%%%%%%%%%%%%%%%%%%%%%%%%
\subsection{Limiting case}
%%%%%%%%%%%%%%%%%%%%%%%%%%%%%%%%%%%%%%

Another interesting case one should discuss is
the resonant tunneling diode made of graphene in the presence of a magnetic field.
This will allow us to characterize the system behavior and underline
their properties. This
limiting situation of the device
is one where the barriers are described by a scalar Lorentz
potential of the form~\cite{peres}
%\begin{eqnarray}
\beq\label{slopo}
V(x,y) = \lim_{\epsilon\rightarrow 0}g \frac
{1}{2\epsilon}[1-\theta(\vert x\vert-\epsilon)]\sigma_z
%+\lim_{\epsilon\rightarrow 0}g \frac
%{1}{2\epsilon}[1-\theta(\vert x-d\vert-\epsilon)]\sigma_z
=g\sigma_z\delta(x). %+\delta(x-d)] \,. \label{pot}
\eeq
The connection with the true barrier is made by identifying $g$
with $2\gamma t'w a$ where $\gamma$ a numerical constant of dimensions inverse
of length and $a$ is the carbon-carbon distance.
This form of
the potential is equivalent to a mass term and therefore to a gap in
the spectrum. However, given the short range nature of the
potential, its effect comes only in the boundary conditions imposed
on
 the wavefunction at the potential position. Note in passing that,
the problem of Dirac electrons in delta function potentials has been
studied in the past~\cite{McKellar87a} and for a recent
review one can consult~\cite{peres}.

To study (\ref{slopo}) in terms of our language, we adopt
the same method as has been applied to the diode case for
zero field~\cite{peres}.
It is based on
the boundary condition around the point $x=0$ of the eigenspinors. This is
equivalent to write
\begin{equation}\label{bceq}
\left(
\begin{array}{c}
\varphi_{1}(0^{+}) \\
 \varphi_{2}(0^{+})\end{array}\right)= M\left(
\begin{array}{c}
\varphi_{1}(0^{-})\\
\varphi_{2}(0^{-})\end{array}\right)
\end{equation}
where the matrix $M$ is given by
\beq
M= \left(
% use packages: array
\begin{array}{ll}
\cosh$(\~{g})$& i\sinh$(\~{g})$ \\
-i\sinh$(\~{g})$ & \cosh$(\~{g})$
\end{array}\right).
\eeq
The points $0^{\pm}$ represent positive and negative
infinitesimals, the constant is $\tilde{g} =\frac{2\gamma t'\omega a}{v_{F}\hbar} $.
For some reasons,
we need the inverse of (\ref{bceq}), which can be obtained
by determining $M^{-1}$. It is easy to get
\begin{equation}
M^{-1}=\ \left(
% use packages: array
\begin{array}{ll}
\cosh$(\~{g})$& -i\sinh$(\~{g})$ \\
i\sinh$(\~{g})$ & \cosh$(\~{g})$
\end{array}\right)
\end{equation}
which together they verify the unitary condition
$M^{-1}M={\mathbb I}_{2}$. Combining all to end up with
the inverse of (\ref{bceq}), such as
\begin{equation}
\left(
\begin{array}{c}
\varphi_{1}(0^{-})\\
\varphi_{2}(0^{-})\end{array}\right)=M^{-1}\left(
\begin{array}{c}
\varphi_{1}(0^{+}) \\
 \varphi_{2}(0^{+})\end{array}\right).
\end{equation}
After getting these two equations, one can use the former analysis
to show that how the obtained results for diode in magnetic field will
be affected. For this, we consider also
two cases: propagation with positive and negative incidences.

%%%%%%%%%%%%%%%%%%%%%%%%%%%%%%%%%%%%%%%%%%%%%%%%%%%%%%%%%%%%%%%%%%%%
{\bf Propagation with positive incidence:}
%%%%%%%%%%%%%%%%%%%%%%%%%%%%%%%%%%%%%%%%%%%%%%%%%%%%%%%%%%%%%%%%%%%%
Returning to our tools, we can map (\ref{bceq}) in terms
of the  cylindrical parabolic functions as
\begin{equation}
\left(
\begin{array}{c}
\varphi_{1}(0^{-})\\
\varphi_{2}(0^{-})\end{array}\right)= \left(
% use packages: array
\begin{array}{ll}
-siD_{|n|-1}(x_{0})& si(-1)^{|n|}D_{|n|-1}(x_{0}) \\
D_{|n|}(x_{0})& (-1)^{|n|}D_{|n|}(x_{0})
\end{array}\right)
\left(
\begin{array}{c}
A^{+}\\
B^{+}\end{array}\right)
\end{equation}
and in similar way, we have
\begin{equation}
\left(
\begin{array}{c}
\varphi_{1}(0^{+})\\
\varphi_{2}(0^{+})\end{array}\right)= C^{+}\left(
\begin{array}{c}
-siD_{|n|-1}(x_{0})\\
D_{|n|}(x_{0})\end{array}\right)
\end{equation}
where $A^+, B^+$ and $C^+$ are three parameters those
should be determined.
According to (\ref{bceq}), the last equation
takes the form
\begin{equation}
\left(
\begin{array}{c}
\varphi_{1}(0^{+})\\
\varphi_{2}(0^{+})\end{array}\right)= M\left(
% use packages: array
\begin{array}{ll}
-siD_{|n|-1}(x_{0})& si(-1)^{|n|}D_{|n|-1}(x_{0}) \\
D_{|n|}(x_{0})& (-1)^{|n|}D_{|n|}(x_{0})
\end{array}\right)
\left(
\begin{array}{c}
A^{+}\\
B^{+}\end{array}\right).
\end{equation}
%\begin{equation}
%= \left(
% use packages: array
%\begin{array}{ll}
%-si\cosh$(\~{g})$D_{|n|-1}(x_{0})+i\sinh$(\~{g})$D_{|n|(x_{0})}& si(-1)^{|n|}\cosh$(\~{g})$D_{|n|-1}(x_{0})+ %i(-1)^{|n|}\sinh$(\~{g})$D_{|n|}(x_{0})\\
%-s\sinh$(\~{g})$D_{|n|-1}(x_{0})+\cosh$(\~{g})$D_{|n|}(x_{0})& s(-1)^{|n|}\sinh$(\~{g})$D_{|n|-1}(x_{0})+
%(-1)^{|n|}\cosh$(\~{g})$D_{|n|}(x_{0})
%\end{array}\right)
%\left(
%\begin{array}{c}
%A^{+}\\
%B^{+}\end{array}\right)
%\end{equation}
This can be solved to obtain $A^{+}$ and $B^{+}$ as
%\begin{equation}
\beqar
A^{+}&=&sC^{+}\, \frac{2s\cosh(\tilde{g})D_{|n|}(x_{0})D_{|n|-1}(x_{0})+
\sinh(\tilde{g}) \left[D_{|n|}^{2}(x_{0})+D_{|n|-1}^{2}(x_{0})\right]}
{2D_{|n|}(x_{0})D_{|n|-1}(x_{0})}\\
%\end{equation}
%and
%\begin{equation}
B^{+}&=&sC^{+}(-1)^{|n|}\, \frac{\sinh(\tilde{g})\left[D_{|n|-1}^{2}(x_{0})-D_{|n|}^{2}(x_{0})\right]}
{2D_{|n|}(x_{0})D_{|n|-1}(x_{0})}.
%\end{equation}
\eeqar
The usefulness of these is to define the reflexion and transmission coefficients
for the present situation. Indeed,
After calculation, we show
%\begin{equation}
\beqar
r^{+} &=& \frac{B^{+}}{A^{+}}=\frac{(-1)^{|n|}\sinh(\tilde{g})
\left[D_{|n|-1}^{2}(x_{0})-D_{|n|}^{2}(x_{0})\right]}
{2s\cosh(\tilde{g})D_{|n|}(x_{0})D_{|n|-1}(x_{0})+\sinh(\tilde{g})
\left[D_{|n|}^{2}(x_{0})+D_{|n|-1}^{2}(x_{0})\right]}\\
%\end{equation}
%\begin{equation}
t^{+} &=& \frac{C^{+}}{A^{+}}=\frac{2sD_{|n|}(x_{0})D_{|n|-1}(x_{0})}
{2s\cosh(\tilde{g})D_{|n|}(x_{0})D_{|n|-1}(x_{0})+
\sinh(\tilde{g})\left[D_{|n|}^{2}(x_{0})+D_{|n|-1}^{2}(x_{0})\right]}.
%\end{equation}
\eeqar

%%%%%%%%%%%%%%%%%%%%%%%%%%%%%%%%%%%%%%%%%%%%%%%%%%%%%%%%%%%%%%%%%%%%
{\bf Propagation with negative incidence:}
%%%%%%%%%%%%%%%%%%%%%%%%%%%%%%%%%%%%%%%%%%%%%%%%%%%%%%%%%%%%%%%%%%%%
In this case, we can write for $0^{+}$
\begin{equation}
\left(
\begin{array}{c}
\varphi_{1}(0^{+})\\
\varphi_{2}(0^{+})\end{array}\right)= \left(
% use packages: array
\begin{array}{ll}
-siD_{|n|-1}(x_{0})& si(-1)^{|n|}D_{|n|-1}(x_{0}) \\
D_{|n|}(x_{0})& (-1)^{|n|}D_{|n|}(x_{0})
\end{array}\right)
\left(
\begin{array}{c}
A^{-}\\
B^{-}\end{array}\right)
\end{equation}
as well as for $0^{-}$
\begin{equation}
\left(
\begin{array}{c}
\varphi_{1}(0^{-})\\
\varphi_{2}(0^{-})\end{array}\right)= C^{-}\left(
\begin{array}{c}
-siD_{|n|-1}(x_{0})\\
D_{|n|}(x_{0})\end{array}\right).
\end{equation}
%\begin{equation}
%\left(
%\begin{array}{c}
%\varphi_{1}(0^{-})\\
%\varphi_{2}(0^{-})\end{array}\right)= M^{-}\left(
%\begin{array}{c}
%\varphi_{1}(0^{+})\\
%\varphi_{2}(0^{+})\end{array}\right)
%\end{equation}
Using the same technique as above, one can obtain
\beqar
r^{-} &=&\frac{B^{-}}{A^{-}}=\frac{(-1)^{|n|}
\sinh(\tilde{g})\left[D_{|n|}^{2}(x_{0})-D_{|n|-1}^{2}(x_{0})\right]}
{2s\cosh(\tilde{g})D_{|n|}(x_{0})D_{|n|-1}(x_{0})-
\sinh(\tilde{g})\left[ D_{|n|}^{2}(x_{0})+D_{|n|-1}^{2}(x_{0})\right]}\\
%\end{equation}
%\begin{equation}
t^{-}&=&\frac{C^{-}}{A^{-}}=\frac{2sD_{|n|}(x_{0})D_{|n|-1}(x_{0})}
{2s\cosh(\tilde{g})D_{|n|}(x_{0})D_{|n|-1}(x_{0})-\sinh(\tilde{g})
\left[D_{|n|}^{2}(x_{0})+D_{|n|-1}^{2}(x_{0})\right]}.
%\end{equation}
\eeqar

Having all ingredients, let us check the probability condition.
This can be achieved by defining two quantities
in terms of the above coefficients, such as
\beqar
R&=&r^{+}r^{-}=\frac{-\sinh^{2}(\tilde{g})\left[D_{|n|}^{2}(x_{0})-D_{|n|-1}^{2}(x_{0})\right]^{2}}
{4\cosh^{2}(\tilde{g})D_{|n|}^{2}(x_{0})D_{|n|-1}^{2}(x_{0})-
\sinh^{2}(\tilde{g})\left[(D_{|n|}^{2}(x_{0})+D_{|n|-1}^{2}(x_{0})\right]^{2}} \label{rR}\\
T&=&t^{+}t^{-}=\frac{4D_{|n|}^{2}(x_{0})D_{|n|-1}^{2}(x_{0})}
{4\cosh^{2}(\tilde{g})D_{|n|}^{2}(x_{0})D_{|n|-1}^{2}(x_{0})-
\sinh^{2}(\tilde{g})\left[D_{|n|}^{2}(x_{0})+D_{|n|-1}^{2}(x_{0})\right]^{2}}\label{rT}.
\eeqar
After a straightforward calculation, one can find a
probability one
\begin{equation}
R+T=\frac{4D_{|n|}^{2}(x_{0})D_{|n|-1}^{2}(x_{0})-
\sinh^{2}(\tilde{g})\left[D_{|n|}^{2}(x_{0})-D_{|n|-1}^{2}(x_{0})\right]^{2}}
{4\cosh^{2}(\tilde{g})D_{|n|}^{2}(x_{0})D_{|n|-1}^{2}(x_{0})-
\sinh^{2}(\tilde{g})\left[D_{|n|}^{2}(x_{0})+D_{|n|-1}^{2}(x_{0})\right]^{2}}=1.
\end{equation}
This also has been obtained by studying the scalar Lorentz potential
in zero field~\cite{peres}.

Let us analyze some limits of the transmission coefficient. This can be reached by
inspecting (\ref{rR}) and as well as (\ref{rT}). Clearly, there is a trivial
solution, i.e. $T=1$, which can easily be obtained by requiring that
the involved parameter $\tilde g$ is nothing but zero.
Furthermore, other interesting discussions can be reported by looking some
cases involving the parameter $g$.
After calculation, we end up with the results

\begin{itemize}
\item   $ g\ll  v_{F}\hbar$ \,  $\Longrightarrow$  $T \ \lga 1$.
\item
  $ g\gg  v_{F}\hbar$ \,  $\Longrightarrow$   $T \ \lga 0$.
\end{itemize}
Of course, one can also try to plot different figures in order to
give a full description of the transmission behavior with respect to
the scalar Lorentz potential taken here.

%%%%%%%%%%%%%%%%%%%%%%%%%%%%%%%%%%%%%%%
\section{Conclusion}
%%%%%%%%%%%%%%%%%%%%%%%%%%%%%%%%%%%%%%

The present paper is devoted to the study of the
tunneling effect for Dirac fermions in
the presence of a constant magnetic field $B$.
More precisely, we started by splitting our system
in two different regions where the second one is
characterized by an energy gap $t'$. Moreover,
we derived the energy spectrum and the corresponding
eigenspinors for each region.
They are used to deal with our task and in particular show
how Dirac fermions behave in such system.
From the energy conservation $E$, we established an interesting
relation between two quantum numbers $(n,m)$ involved in game and
the system energy. This allowed us to control the behavior
of the energy and therefore fix the allowed values
of $(n,m)$ where their behaviors are plotted in
the  figures $(1,2,3)$. In analyzing their properties,
we noticed that there are three limiting cases  those
have significant influences on the study of
the tunneling effect.

Subsequently, we concentrated on two examples of the present system. Indeed,
by considering a barrier in magnetic field, we analyzed the propagation
for positive and negative incidences. In fact, using the continuity equations
we determined the corresponding reflexion and transmission coefficients.
They are obtained as a function of the ratio ${E\over t'}$ and more
importantly, they verified the probability condition, i.e.
 their sum is equal one. To characterize transmission
of the system, we plotted
the figures (4,5,6) showing its behaviors for different values
of the involved parameters. Using the three limiting cases of $(n,m)$,
we noticed that the system can be seen either as a diopter or
mirror depending to which case is included. These actually
are corresponding  to a total reflexion or total transmission,
respectively.

We focused on another interesting case that is a diode
in magnetic field.
This is equivalent to consider a system composed
of
three regions where the second regions is completely different
from the first and third, which they are identical.
After getting different coefficients for positive and
negative incidences, we defined two new quantities to
end up with their sum is equal one, i.e. probability one. Requiring some constraint
on the system, we showed that it is possible to have a total transmission.
 Moreover, we considered the case where the barriers
are described by a scalar Lorentz potential of the form
given in~(\ref{slopo}). This allowed
us to study the resonant tunneling of diode make in graphene
where different coefficients are obtained and their
probability condition is verified. Moreover, we showed that
there is a possibility to have a total transmission in such case.

Finally, our findings so far will not remain at this stage. In fact,
still some interesting questions one should answer to get a full
descriptions of different issues.
%we
%project to deal with and discuss other issues.
These concern for instance the Fabry--P\'erot solid etalon
in the usual optics~\cite{BornWolf} and
the related finesse factor. Since we have transmission,
one can also ask about the conductivity
to discuss the anomalous quantum Hall effect
for the present system.
Hopefully to come back to these
questions in a subsequent work.

%%%%%%%%%%%%%%%%%%%%%%%%%%%%%%%
\section*{Acknowledgment}
%%%%%%%%%%%%%%%%%%%%%%%%%%%%%%%

This work was completed during a visit of AJ to the Abdus Salam Centre for Theoretical
Physics (Trieste, Italy) in the framework of junior associate scheme.
He would like to acknowledge the
financial support of the centre.

%%%%%%%%%%%%%%%%%%%%%%%%%%%%%%%%%%%%%%%%%%%%%%%%%%

\end{document}